\documentstyle[prl,aps,multicol]{revtex}
\tightenlines
\newcommand{\beq}{\begin{equation}}
\newcommand{\eeq}{\end{equation}}
\newcommand{\bea}{\begin{eqnarray}}
\newcommand{\eea}{\end{eqnarray}}
\begin{document}

\title{Helium mixtures in nanotube bundles}

\author {George Stan$^{1,2,4}$, Jacob M. Hartman$^1$, Vincent H. Crespi$^1$,
Silvina M. Gatica$^{1,3}$,Milton W. Cole$^{1}$}
\address{ $^1$Department of Physics, Penn State University, University
Park, PA 16802, USA \\
$^2$Present address: Institute for Physical Science $\&$ Technology and\\
Department of Chemical Engineering,
University of Maryland, College Park, MD 20742, USA\\ 
$^3$Departamento de F\'{\i}sica, Facultad de Ciencias Exactas y Naturales,\\
Universidad de Buenos Aires, Buenos Aires 1428, Argentina\\
$^4${corresponding author, e-mail: stan@phys.psu.edu}}

\maketitle
\begin{abstract}
An analogue to Raoult's law is determined for the case of a $^3$He\,--$^4$He
mixture adsorbed in the interstitial channels of a bundle of carbon
nanotubes. Unlike the case of He mixtures in other environments, the ratio
of the partial pressures of the coexisting vapor is found to be a simple 
function of the ratio of concentrations within the nanotube bundle. 

\end{abstract}

\baselineskip 14pt
\vskip.5cm

Helium atoms are strongly attracted to and absorbed within nanotube bundles
\cite{stan1,stan3,teizer}. For tubes at the experimentally observed
diameter of $\sim$14 Angstroms, the most energetically favorable sites lie
within interstitial channels bounded by three nanotubes; these tubes form a
hexagonal array. For a system of adsorbed $^4$He atoms, the
individual atoms are highly localized by the periodic potential due to the
surrounding nanotubes\cite{anisotropic}. The He atoms' mutual interactions
induces a condensation which is well-described by an anisotropic lattice
gas model. Because of the well-localized atomic states, the transition
temperature to the condensed state is nearly the same for $^3$He and
$^4$He. Here we analyze a mixture of these isotopes and show that the
system forms an ideal solution, wherein the ratio of the partial pressures
of the coexisting vapors of the two components satisfies an analogue of
Raoult's law\cite{raoult}.

Within the grand canonical ensemble, the term contributed to the partition
function by any specific quantities N$_3$ and N$_4$ of $^{3}$He and
$^{4}$He is\cite{spin}
\beq
p(N_3, N_4)\propto \exp ( \beta [N_3(\mu_3 - \epsilon_3) + N_4(\mu_4 -
\epsilon_4)]) {N! \over N_3!\ N_4!} Z_I\,.
\eeq
Here $N = N_3 + N_4 \equiv \theta N_s$, $\theta$ is the total occupation
fraction of the adsorption sites. $N_s$ is the number of adsorption sites,
$\epsilon_{3,4}$ is the single particle energy in the interstitial channel,
$\beta = 1/(k_B T)$ is the inverse temperature, $\mu_{3,4}$ is the chemical
potential of $^{3,4}$He, and $Z_I$ is the canonical partition function of
$N$ indistinguishable He atoms (omitting the single-particle energy). The
energy $\epsilon_{3,4}$ assumes a single value for each species because the
lowest He bands are very narrow ($\sim$0.2 K) and well-separated ($\sim$
100 K) from higher bands\cite{jake}. We can write 
\beq
Z_I = \exp\ (- \beta N f_I)\,,
\eeq
where $f_I (\theta)$ is the Helmholtz free energy per atom for such a system
of interacting indistinguishable particles. We need not specify the form of
$f_I (\theta)$ here; its behavior was described in previous
work\cite{anisotropic} using an anisotropic lattice gas model (based in
turn on studies by Fisher and Graim and Landau\cite{ising}).

The equilibrium number of $^3$He atoms, $<N_3>$, follows from maximizing
$p(N_3, N_4)$:
\beq
\left[{{\partial\ \ln\ p (N_3, N_4)}\over {\partial
N_3}}\right]_{T,\mu_{3,4}, N_4} = 0\,. 
\eeq
Hence we obtain the condition for the chemical potential: 
\bea
\mu_3 &=& \epsilon_3 + \beta^{-1} \ln\ x + f_I (\theta) + \theta\
f_I'(\theta)\nonumber\\ &=& \epsilon_3 + \beta^{-1} \ln\ x + g'(\theta)
\label{eq:mu_3}\\ g(\theta)&\equiv& \theta f_I (\theta)\nonumber\,, 
\eea
where $x=N_3/(N_3+N_4)$ is the $^3$He concentration and a prime refers to
differentiation with respect to $\theta$. The coexisting three dimensional
vapor (assumed ideal) satisfies\cite{spin} 
\beq
\mu_3 = \beta^{-1}\ \ln (n_3 \lambda_3^3)= \beta^{-1}\ \ln (\beta P_3
\lambda_3^3)\,,
\eeq
where $n_3$ is the $^3$He particle density in the vapor phase, $P_3$ is
the $^3$He partial pressure, and $\lambda_3 = (2 \pi \hbar^2 \beta/m_3)^{1/2}$ 
is the de Broglie wavelength for the $^3$He atoms. The resulting isotherm is
then
\beq
{P_3 \lambda_3^3} = x\ \exp [\beta (\epsilon_3 + g'(\theta))]\,. 
\eeq
In similar fashion,
\beq
P_4 \lambda_4^3 = (1-x) \exp [\beta (\epsilon_4 + g'(\theta))]\,, 
\eeq
which yields a remarkably simple relation between the isotopic partial
pressures:
\beq
{P_3 \over P_4} = {\left(3 \over 4\right)}^{3/2}\ {x \over 1 -x}\ e^{\beta
(\epsilon_3 - \epsilon_4)}\,.
\label{eq:raoult}
\eeq
The ratio of partial pressures is independent of both $\theta$ and the form
of the interaction between the atoms. These quantities disappear from the
pressure ratio since the interaction between He atoms is nearly
isotope-independent due to the strong atomic localization. This ratio is an
expression of Raoult's law of solutions. Because $\epsilon_3-\epsilon_4
\simeq 17$ K, $P_3 > P_4$ except at small $x$ and high $T$.

This result coincides with that obtained from a noninteracting (band)
model. In this case the average number of $^4$He particles is \beq
N_4 = \int d\epsilon {{\cal{N}} (\epsilon) \over e^{\beta (\epsilon -
\mu_4)} -1} \eeq
where ${\cal{N}} (\epsilon)$ is the density of states as a function of the
single-particle energy $\epsilon$. In the limiting case of very low
coverage relevant to this noninteracting model (i.e.\ $exp(- \beta \mu) >>
1$), 
\beq
N_4 \simeq e^{\beta \mu_4} \int d\epsilon {\cal{N}} (\epsilon) e^{- \beta
\epsilon}\,.
\label{eq:N_4}
\eeq
For the present case of a very narrow band we can approximate the density
of states by a delta function to obtain
\beq
N_4 \simeq e^{\beta (\mu_4 - \epsilon_4) }\ {L \over a}\,,
\eeq
where $a$ is the lattice constant and $L$ is the total length of
interstitial channel. The chemical potential is then:
\beq
\mu_4 = \epsilon_4 + \beta^{-1}\ \ln\ \left({N_4 a \over L}\right)\,.
\label{eq:mu_4_Henry}
\eeq
Taking the same vapor chemical potential as before, and following a similar
analysis for $^3$He, one again obtains Eqn. (\ref{eq:raoult}) as the
isotopic ratio of the partial pressures. This confirms our expectation that
the pressure ratio in a localized lattice gas model with arbitrary
interactions coincides with that obtained in the appropriate noninteracting
band model. At high density the noninteracting model fails, so a
comparison is not appropriate.

These results depend upon the isotope-independence of the He\,--He
interaction, a property which we now address in detail. Three effects could
contribute to an isotope dependence in the He\,--He interaction: differences
in zero point motion (primarily along the axis of the channel), the
magnetic interaction in $^3$He, and the exchange interaction. The
contribution from zero point motion can be described by a Hartree
interaction,
\beq
V_H(a)=\int d\vec r_1\; n_1(\vec r_1) \int d\vec r_2 \;n_2(\vec r_2)\;
u(r_{12})\,,
\eeq
where $n_1(\vec r_1)$ and $n_2(\vec n_2)$ are the densities at neighboring
sites separated by a distance $a$ in the same channel, and $u(r_{12})$ is
the interaction potential between two atoms. Since the densities are
well-confined, we can Taylor expand the integrand (with $z_{1,2}$ the axial
coordinate and $\rho_{1,2}$ the transverse radial coordinate). Keeping
second order terms, we obtain\cite{hartree}
\beq
V_H(a)=u(a)+\frac{1}{2a} u'(a)(<\rho^2_1>+<\rho^2_2>)+\frac{1}{2}u''(a)
(<z^2_1>+<z^2_2>)\,.
\eeq
Using the single-particle wavefunctions for $^3$He and $^4$He in the
interstitial channel of an (18,0)\cite{indices} tube lattice\cite{jake}, we
obtain a correction $\Delta V \equiv V_H(a) - u(a)$ of -0.488 K, -0.490 K
and -0.489 K respectively for the $^4$He\,--$^4$He, $^3$He\,--$^3$He and
$^3$He\,--$^4$He interactions. Although the magnitude of this correction
reaches 25\% of the bare He\,--He interaction, it introduces only a tiny
distinction between the isotopes. 

The magnetic energy of $^3$He is also negligible ($\sim$ nK) on this scale
due to the very small nuclear moment of $^3$He. As to the exchange of He
atoms between different sites, one might expect this to be significantly
different for the two species. However, the very similar effective masses
and band widths for the two isotopes\cite{jake} suggests that the effects
of exchange are not significantly different for the two species.

In summary, we have found a simple expression for the ratio of partial
pressures of the ambient vapor of He isotopes when exposed to adsorption
sites within bundles of carbon nanotubes. To our knowledge, this expression
has no precedent in the field of $^3$He\,--$^4$He mixtures\cite{ramos}. In 
other known adsorption situations, the single-particle wave functions are not
sufficiently localized to justify the present assumption that the effects
of interactions are nearly isotope-independent. One can contrast the localized
wave functions in the present case\cite{jake} with those of He on
graphite\cite{carlos1} wherein the states are quite delocalized. The band
masses are $m^*/m \sim 18$ in the nanotube environment and $\sim 1.05$ on
graphite \cite{carlos2}. In the case of He on graphite one cannot employ
the present simple analysis because the effect of interparticle
interactions depends on the species degree of localization, which differs
for the two isotopes\cite{carlos2}. It is plausible that other nanoscale
porous media, such as zeolites, exhibit similar behavior of adsorbed He
isotopic mixtures.

We thank V. Bakaev, and W. A. Steele for helpful discussions. This research
has been supported by the Army Research Office, the National Science
Foundation under grant DMR-9876232, and the donors of the Petroleum
Research Fund, administered by the American Chemical Society.


\begin{references}

\bibitem{stan1}
G. Stan and M. W. Cole, Surf. Sci. {\bf 395}, 280 (1998).

\bibitem{stan3}
G. Stan, M. Boninsegni, V. H. Crespi, and M. W. Cole, J. Low Temp. Phys.
{\bf 113}, 447 (1998), cond-mat/9808229.

\bibitem{teizer}
W. Teizer, R. B. Hallock, E. Dujardin, and T. W. Ebbesen, Phys. Rev. Lett.
{\bf 82}, 5305 (1999).

\bibitem{anisotropic}
M. W. Cole, V. H. Crespi, G. Stan, J. M. Hartman, S. Moroni, and M.
Boninsegni, submitted to Phys. Rev. Lett., cond-mat/9908234.

\bibitem{raoult}
See, for example, G. N. Lewis, {\it Thermodynamics}, McGraw-Hill, New York,
(1961).

\bibitem{spin}
For simplicity, we consider spinless particles throughout this paper. Final
results, however, apply to the general case of atoms with spin.

\bibitem{jake}J. M. Hartman and V. H. Crespi, unpublished.

\bibitem{ising}
M. E. Fisher, Phys.Rev. {\bf 162}, 480 (1967); T. Graim and D. P. Landau,
Phys. Rev. B {\bf 24}, 5156 (1981).

\bibitem{hartree}
The reader recognizes that the Hartree interaction is divergent because of
the omission of the pair correlation function; this spurious divergence
disappears in the present expansion.

\bibitem{indices}
Nanotubes are labelled according to the lattice coordinates of the
circumference vector. See, for example, R. Saito, M. Fujita, G.
Dresselhaus, and M. S. Dresselhaus, Appl. Phys. Lett. {\bf 60}, 2204
(1992).

\bibitem{ramos}
For recent studies of helium mixtures on graphite see R. C. Ramos, 
P. S. Ebey, and O. E. Vilches, J. Low Temp. Phys. {\bf 110}, 615 (1998) and 
R. C. Ramos and O. E. Vilches, J. Low Temp. Phys. {\bf 113}, 981 (1998).

\bibitem{carlos1}
W. E. Carlos and M. W. Cole, Surf. Sci. {\bf 91}, 339 (1980).

\bibitem{carlos2}
W. E. Carlos and M. W. Cole, Phys. Rev. B {\bf 31}, 3713 (1980).

\end{references}
\end{document}